\documentclass[a4paper,11pt]{article}
\usepackage{pos}

\title{A new puzzle in non-leptonic B decays}
\author*[a]{Aritra Biswas}
\author[b]{S\'ebastien Descotes-Genon}
\author[a]{Joaquim Matias}
\author[b,c]{Gilberto Tetlalmatzi-Xolocotzi}
\affiliation[a]{Universitat Aut\`onoma de Barcelona, 08193 Bellaterra, Barcelona,\\
	Institut de F\'{i}sica d'Altes Energies (IFAE), The Barcelona Institute of Science and Technology, Campus UAB, 08193 Bellaterra (Barcelona)}
\affiliation[b]{Universit\'e Paris-Saclay, CNRS/IN2P3, IJCLab, 91405 Orsay, France}
\affiliation[c]{Theoretische Physik 1, Center for Particle Physics Siegen (CPPS), Universit\"at Siegen, Walter-Flex-Str. 3, 57068 Siegen, Germany}

\emailAdd{abiswas@ifae.es}
\emailAdd{sebastien.descotes-genon@ijclab.in2p3.fr}
\emailAdd{matias@ifae.es}
\emailAdd{gtx@physik.uni-siegen.de}

\abstract{We propose a set of new optimized observables using penguin mediated $\bar{B}_d$ and $\bar{B}_s$ decays:
	${\bar B}_{d,s} \to K^{*0} \bar{K}^{*0}$, ${\bar B}_{d,s} \to K^{0} \bar{K}^{0}$, ${\bar B}_{d,s} \to K^{0} \bar{K}^{*0}$ and ${\bar B}_{d,s} \to \bar{K}^{0} {K^{*0}}$ together with their CP conjugate partners. These observables are substantially cleaner than the corresponding branching ratios, which are plagued by large end point divergences. We find that the dominant contribution to the uncertainties of these observables stem from the corresponding form factors. The Standard Model estimates for these observables corresponding to the $K^{*0}\bar{K}^{*0}$ and $K^0\bar{K}^0$ final states are in tension with their respective experimental numbers at the $\sim2.5 \sigma$ level. The pattern of deviations w.r.t these observables as well as the individual branching ratios suggest that a possible explanation might be new physics both in $b\to s$ and $b\to d$ transitions. We find that, taken one at a time, only the Wilson coefficients $C_{4d,s}^{NP}$ and $C_{8gd,s}^{NP}$ can potentially satisfy all the current experimental data on the branching ratios as well as the optimized observables. Furthermore, such observables involving mixed (pseudoscalar-vector) states like $K^{*0}\bar{K}^0$ etc show distinctive patterns sensitive to these different new physics explanations.}

\FullConference{21st Conference on Flavor Physics and CP Violation (FPCP 2023)\\
 29 May - 2 June 2023\\
 Lyon, France\\}

\begin{document}
	\maketitle

	\section{Introduction}
	In a recent article~\cite{Alguero:2020xca} the authors studied the observable $L_{{K}^* \bar{K}^*}$ defined as the ratio of the longitudinal branching ratios of $\bar{B}_s \to K^{*0} \bar{K}^{*0}$ versus $\bar{B}_d \to K^{*0} \bar{K}^{*0}$. This observable exhibits a tension of 2.6$\sigma$ between its Standard Model (SM) prediction and data. The authors calculated the SM estimate under various frameworks, viz-a-viz: $SU(3)$ symmetry (see Ref.~\cite{Amhis:2022hpm} for a recent example of this type of analysis combining $\bar{B}_{s}\to K^{0}\bar{K}^{0}$ with other isospin and U-spin related modes, Ref.~\cite{Li:2022mtc} for PQCD and Ref.~\cite{Fleischer:1999zi,Fleischer:1999pa,Fleischer:2007hj,Fleischer:2010ib,Fleischer:2016jbf,Bhattacharya:2022akr} for other attempts), but also QCD Factorisation (QCDF)~\cite{Beneke:2000ry,Beneke:2001ev,Beneke:2003zv,Beneke:2006hg,Bartsch:2008ps}, or even a combination of the two approaches~\cite{Descotes-Genon:2006spp,Descotes-Genon:2007iri,Descotes-Genon:2011rgs,Alguero:2020xca}. Having performed a model-independent analysis of this tension within the Weak Effective Theory at the $b$-quark mass scale (keeping the analysis at the level of the operators generated in the SM and their chirally-flipped counterparts), they ended up with an 
	explanation relying on New Physics (NP) contributions to the Wilson coefficients of two operators: a)  the QCD penguin operator $O_{4s}=(\bar{b}_i s_j)_{V-A} \sum_q (\bar{q}_j q_i)_{V-A}$ and b) the chromomagnetic operator $O_{8g}=- \frac{g_s}{8\pi^2} m_b \bar{s} \sigma_{\mu\nu} (1+\gamma_5) G^{\mu\nu}b$.  
	
	In order to further understand this deviation and confirm its NP origin, one has to:
	\begin{itemize}
		\item identify other modes which exhibhit a similar sensitivity to the same NP,
		\item design observables for these modes with a reduced sensitivity to hadronic uncertainties, 
		\item make sure these additional observables yield clear patterns of consistent deviations for different modes depending on the NP scenario considered. 
	\end{itemize}
	Under these conditions, one should be able to confirm the NP origin of the deviations observed with reasonable clarity. These observables can be complemented by additional less clean observables that favour one scenario with respect to another in a more qualitative way.
	
	Keeping the above set of points in mind, we extend the discussion to a larger set of 
	decays, with the same underlying quark transitions and thus a similar potential sensitivity to NP, but with different 
	final states in the scope of this article. In practice we consider non-leptonic $\bar{B}_d$ and $\bar{B}_s$ meson decays not only to two vectors ($VV$), but also two pseudoscalars ($PP$) and a vector and pseudoscalar ($PV$ or $VP$), with $V=K^{*0}$ and $P=K^0$. We identify and construct observables with reduced hadronic uncertainties following the same strategy as for $L_{K^*\bar{K}^*}$ discussed in ref.~\cite{Alguero:2020xca}. Some of the simple NP scenarios suggested to explain 
	$L_{K^*\bar{K}^*}$ yield distinct and consistent patterns of deviations for the other modes. The additional observables for these other modes have very similar values in the SM but they can differ by one order of magnitude in some NP scenarios. Such hierarchy among observables, which can hardly be attributed to residual hadronic uncertainties, may be 
	tested at LHCb and Belle II. Interestingly, the individual branching ratios show patterns of deviations that lead us towards investigating the possibility of NP affecting both $b\to d$ and $b\to s$ transitions. 
	
	The structure of the proceeding is as follows. In section~\ref{sec:th} we introduce and discuss the theoretical framework that will be used for the construction of the observables in sections~\ref{sec:th-VV},~\ref{sec:th-PP} and~\ref{sec:th-PV-VP}.  Some of the observables designed in this section will require an LHCb upgrade to be accessible. We provide the SM and experimental estimates for these observables in section~\ref{sec:obs_val_SM_exp}. A model independent analysis first assuming NP only in $b\to s$ and then in both $b\to s,d$ transitions as a potential explanation for both the $L_{K^{(*)}K^{(*)}}$ has been discussed in section~\ref{sec:obs_val_SM_exp}. In section~\ref{sec:pattern} we discuss the role of the observables with mixed finas states in disentangling the dominant NP contributions when data on these modes is available in the future. We draw our conclusions in section~\ref{sec:conclusions}.
	
	\section{Theory}\label{sec:th}
	\begin{figure} 
		\begin{center}
			\includegraphics[width=6.5 cm]{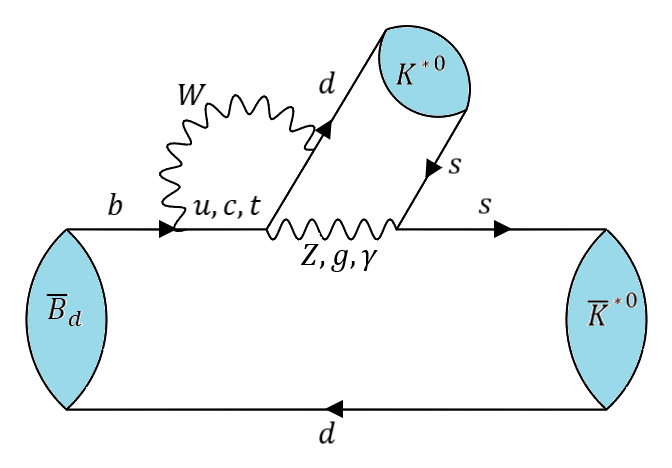} \qquad
			\includegraphics[width=6.5 cm]{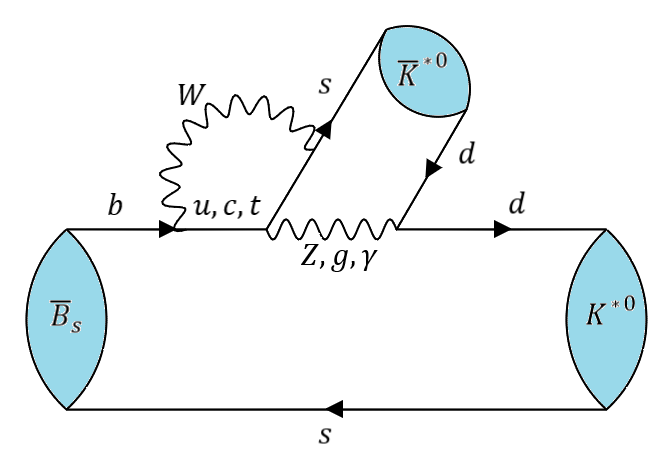}
		\end{center}  \caption{SM contributions to the non-leptonic decays $\bar{B}_{d,s}\to K^{*0} \bar{K}^{*0}$. The internal quark lines consist in a $u$-type quark, the curved wavy line is a $W$ boson, and the horizontal wavy line may be a gluon, a photon or a $Z$ boson, leading to different types of penguins. }
		\label{fig:kstarkstar}   
	\end{figure}
	\subsection{$\bar{B}_{d,s}\to K^{*0}\bar{K}^{*0}$ and $L_{K^*\bar{K}^*}$}\label{sec:th-VV}
	As discussed in ref.~\cite{Alguero:2020xca}, the final state for a $\bar{B_q}\to K^{*0}\bar{K}^{*0}$ can be in one of three different polarisation states. One can always write down the corresponding amplitude as:
	\begin{equation}
	\bar{A}_f\equiv A(\bar{B}_q\to K^{*0} \bar{K}^{*0})
	=\lambda_u^{(q)} T_q + \lambda_c^{(q)} P_q
	=\lambda_u^{(q)}\, \Delta_q - \lambda_t^{(q)} P_q 
	\label{dec}
	\end{equation}
	where $\lambda_U^{(q)}=V_{Ub} V_{Uq}^*$~\footnote{The weak phase in $\lambda_t^{(q)}$ is the angle $\beta_q$, defined as
		$\beta_q\equiv \arg \left(- \frac{V_{tb} V_{tq}^*}{V_{cb} V_{cq}^*} \right)= \arg \left(- \frac{\lambda_t^{(q)}}{\lambda_c^{(q)}} \right)\,,
		$} and $\Delta_q=T_q-P_q$. The CP-conjugate amplitude is given by 
	\begin{equation}
	A_{\bar{f}}=(\lambda_u^{(q)})^* T_q + (\lambda_c^{(q)})^* P_q    =(\lambda_u^{(q)})^* \Delta_q - (\lambda_t^{(q)})^* P_q\,.
	\end{equation}
	$A_{\bar{f}}$ is related to $A=A(B_q\to  K^{*0}\bar{K}^{*0})=\eta_f A_{\bar{f}}$
	where $\eta_f$ is the CP-parity of the final state, given for $j=0,||,\perp$ respectively as $1,1,-1$. We would like to remind the reader that for this particular case (and in general), $T_q$ and $P_q$  does not (need not) represent the tree and penguin topologies respectively (fig.~\ref{fig:kstarkstar}); and are simply the hadronic matrix elements accompanying the $\lambda_u^{(q)}$ and $\lambda_c^{(q)}$ CKM factors respectively. These matrix elements can be calculated in the framework of QCD factorization (QCDF) where they are expressed as an expansion in $\alpha_s$ upto $1/m_b$ suppressed terms that entail long distance effects and endpoint divergences. For vector-vector final states, a clear hierarchy exists between the different polarizatoins such that only the longitudinal polarization can be accurately accounted for from within the QCDF framework. Hence, from now on and for the rest of this proceeding, any mention of the vector-vector final states will implicitly assume that they are longitudinally polarized. Furthermore, the quantity $\Delta_q$ which is the difference between $T_q$ and $P_q$ is protected from infra red divergences as has been discussed in refs.~\cite{Descotes-Genon:2006spp,Descotes-Genon:2007iri,Descotes-Genon:2011rgs,Alguero:2020xca}. This quantity is expected to be significantly smaller than  both $T_q$ and $P_q$, as can be seen from appendix A.4 of ref.~\cite{Biswas:2023pyw}.
	
	In light of the above considerations we define:
	\begin{eqnarray}\label{eq:Lgeneraldiscussion}
	L_{K^*\bar{K}^*}&=&\rho(m_{K^{*0}},m_{K^{*0}})\frac{{\cal B}({\bar{B}_s \to K^{*0} {\bar K^{*0}}})}{{\cal B}({\bar{B}_d \to K^{*0} {\bar K^{*0}})}}\frac{ f_L^{B_s}}{ f_L^{B_d}}=\frac{|A_0^s|^2+ |\bar A_0^s|^2}{|A_0^d|^2+ |\bar A_0^d|^2}\,,
	\\
	&=&\kappa \left|\frac{P_s}{P_d}\right|^2 
	\left[\frac{1+\left|\alpha^s\right|^2\left|\frac{\Delta_s}{P_s}\right|^2
		+ 2 {\rm Re} \left( \frac{ \Delta_s}{P_s}\right) {\rm Re}(\alpha^s) 
	}{1+\left|\alpha^d\right|^2\left|\frac{\Delta_d}{P_d}\right|^2
		+ 2 {\rm Re} \left( \frac{ \Delta_d}{P_d}\right) {\rm Re}(\alpha^d)} \right]\,
	\end{eqnarray}
	where $\rho(m_1,m_2)$ stands for the ratio of phase-space factors defined by
	\begin{equation}
	\rho(m_1,m_2)=\frac{\tau_{Bd}}{
		\tau_{Bs}}\frac{m_{B_s}^3}{m_{B_d}^3}\frac{\sqrt{(m_{B_d}^2-(m_1+m_2)^2)(m_{B_d}^2-(m_1-m_2)^2)}}{\sqrt{(m_{B_s}^2-(m_1+m_2)^2)(m_{B_s}^2-(m_1-m_2)^2)}},
	\end{equation} $A^q_0$ stands for the amplitude for a $B_q$ meson decaying into a longitudinally polarised $K^{*0}\bar{K}^{*0}$ pair, and the CKM factors read 
	\begin{eqnarray}
	\kappa&=&\left|\frac{\lambda^s_u+\lambda^s_c}{\lambda^s_u+\lambda^s_c} \right|^2=22.91^{+0.48}_{-0.47}, \nonumber\\
	\alpha^d&=&\frac{\lambda^d_u}{\lambda^d_u+\lambda^d_c}=-0.0135^{+0.0123}_{-0.0124} +0.4176^{+0.0123}_{-0.0124}i, \nonumber\\
	\alpha^s&=&\frac{\lambda^s_u}{\lambda^s_u+\lambda^s_c}=0.0086^{+0.0004}_{-0.0004}-0.0182^{+0.0006}_{-0.0006}i.
	\end{eqnarray}
	As can be seen numerically, $\alpha^d$ is Cabibbo allowed, $\alpha^s$ is Cabibbo-suppressed $O(\lambda^2)$, whereas $\Delta_q/P_q$ is expected to be small. Therefore $L_{K^*\bar{K}^*}$ is directly related to the ratio $|P_s/P_d|$, which can be predicted to a good accuracy within QCDF and is protected by $U$-spin symmetry from uncontrolled $1/m_b$-suppressed long-distance contributions.
	
	\subsection{$\bar{B}_{d,s}\to K^0\bar{K}^0$ and $L_{K\bar{K}}$}\label{sec:th-PP}
	An observable similar to $L_{K^*\bar{K}^*}$ can also be constructed for the pseudoscalar final state $K^0\bar{K}^0$. The chain of logic and theoretical motivation behind constructing such an observable is similar to the case of $L_{K^*\bar{K}^*}$, albeit simpler. This is because in this case the di-mesonic final state being pseudsoscalar, there are no polarizations involved and one can use QCDF for a prediction of the branching ratios (BR's) for the $\bar{B}_{d,s}\to K^0\bar{K}^0$ and the corresponding CP conjugate decays. Fig.~\ref{fig:kstarkstar} remains exactly the same, since the valence quark content of the $K$ and the $K^*$ mesons are the same. The observable $L_{K\bar K}$ is hence defined as:
	\begin{equation}\label{eq:LKtKt}
	L_{K\bar{K}}=\rho(m_{K^0},m_{K^0})\frac{{\cal B}({\bar{B}_s \to K^{0} {\bar K^{0}}})}{{\cal B}({\bar{B}_d \to K^{0} {\bar K^{0}}})} =\frac{|A^s|^2+ |\bar A^s|^2}{|A^d|^2+ |\bar A^d|^2}\,,
	\end{equation}
	
	\subsection{$\bar{B}_{d,s} \to K^{0} \bar{K}^{*0}$, $\bar{B}_{d,s} \to \bar{K}^{0} {K^{*0}}$, $\hat{L}_{K^*}$, $\hat {L}_K$, $L_{K^*}$, $\hat {L}_K$ and $L_{Total}$} \label{sec:th-PV-VP}
	In order to complete the discussion, it is only natural that the above chain of discussion is extended to mixed (pseudoscalar-vector (PV) or vector-pseudoscalar (VP)) final states. One distinguishes between them in terms of which meson among the vector and the pseudoscalar receives contributions from the spectator quark. Such separations, although possible at LHCb, comes at a cost. In order to experimentally identify the final state meson with the spectator quark from the parent $B_{d,s}$, one has to employ tagging. This reduces the number of events significantly~\footnote{Numerically, tagging reduces the statistics by about 1/20~\cite{Fazzini:2018dyq}.}. We hence introduce the optimized L observables for the mixed modes in a step-wise manner:
	\begin{itemize}
		\item Observables that require tagging for both the $B_{s,d}$ modes and are accessible during Run 3 of LHCb. These can be divided into:
		\begin{itemize}
			\item $M_1=K^{*0}$ 
			case. We will denote this observable ${\hat L}_{K^*}$:
			\begin{equation}\label{eq:LKst-def}
			{\hat L}_{{K}^{*}}=\rho(m_{K^0},m_{K^{*0}})\frac{{\cal B}({{\bar B}_s\to {{ K^{*0}}\bar K}^{0})}
			}{{\cal B}({{\bar B}_d \to {\bar K}^{*0} { K^{0}})}} =\frac{|A^s|^2+ |\bar A^s|^2}{|A^d|^2+ |\bar A^d|^2}\,,
			\end{equation} 
			\item $M_1=K^0$ case: We will denote this observable ${\hat L}_K$:
			\begin{equation}\label{eq:LK-def}
			{\hat L}_{K}=\rho(m_{K^0},m_{K^{*0}})\frac{{\cal B}({{\bar B}_s \to 
					{ K^{0}}{\bar K}^{*0} )}}{{\cal B}({{\bar B}_d \to {\bar K}^{0} { K^{*0}})}} =\frac{|A^s|^2+ |\bar A^s|^2}{|A^d|^2+ |\bar A^d|^2}\,,
			\end{equation}
		\end{itemize}
		\item Considering the fact that tagging $B_d$ modes is more difficult than tagging the $B_s$ modes, the next obvious step is to construct optimized observables with untagged $B_d$ but tagged $B_s$ modes. These should be accessible with the current Run 1 and 2 data from LHCb. However, the sensitvity of these observables to NP scenarios will be limited:
		\begin{itemize}
			\item For $M_1=K^{*0}$ 
			\begin{equation}\label{eq:LhatKstar} 
			L_{K^{*}}= 2\,\rho(m_{K^0},m_{K^{*0}})\frac{{\cal B}({{\bar B}_s\to  { K^{*0}}
					{\bar K}^{0}    )}}{{\cal B}({{\bar B}_d \to {\bar K}^{*0} { K^{0}})}
				+{\cal B}({{\bar B}_d \to {\bar K}^{0} { K^{*0}})}
			} 
			=\frac{2 R_d}{1+ R_d}\hat{L}_{K^{*}}\,,\end{equation}
			\item For $M_1=K^0$ case: 
			\begin{equation} \label{eq:LhatK} 
			L_K= 2\, \rho(m_{K^0},m_{K^{*0}})\frac{{\cal B}({{\bar B}_s\to  { K^{0}}
					{\bar K}^{*0}    )}}{{\cal B}({{\bar B}_d \to {\bar K}^{*0} { K^{0}})}
				+{\cal B}({{\bar B}_d \to {\bar K}^{0} { K^{*0}})}
			}     =\frac{2}{1+ R_d}\hat{L}_K\,,
			\end{equation}
			reexpressing them in terms of the optimal observables, using $R_d$:
			\begin{equation}
			R^d=\frac{{\cal B}({{\bar B}_d \to {\bar K}^{*0} {{ K}^{0}})}}{{\cal B}({{\bar B}_d \to {\bar K}^{0} { K^{*0}})}}\,.
			\end{equation}   
		\end{itemize}
		\item Lastly, we consider one final observable accessible in the short term, but with further reduced sensitivity to NP, where neither the $B_d$ nor the $B_s$ modes need tagging:
		\begin{eqnarray}
		{L}_{\rm total} &=& \rho(m_{K^0},m_{K^{*0}}) \left(\frac{{\cal B}({\bar{B}_s \to K^{*0} {\bar K^{0}})}+ {\cal B}({\bar{B}_s \to K^{0} {\bar K^{*0}}})}{{\cal B}({\bar{B}_d \to {\bar K}^{*0} { K^{0}})}+ {\cal B}({\bar{B}_d \to {\bar K}^{0} { K^{*0}})}}\right)\nonumber
		\\&=&
		\frac{L_{K^*}+L_K}{2}=
		\frac{\hat{L}_{K}+ \hat{L}_{K^*} R^d}{1+R^d}
		\end{eqnarray}	
	\end{itemize}

	\section{The SM and Experimental values of the Observables} \label{sec:obs_val_SM_exp}
\begin{figure}
		\begin{center}
			\includegraphics[width=7cm]{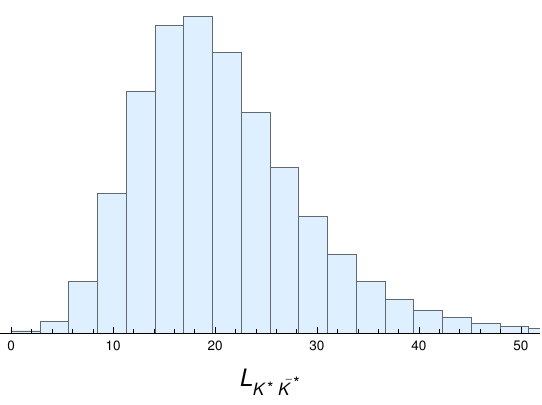}
			\includegraphics[width=7cm]{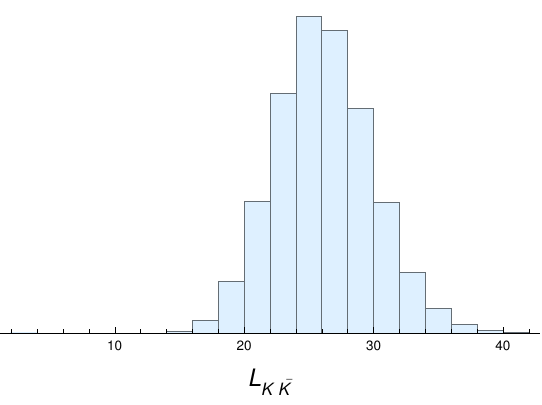}
			\includegraphics[width=7cm]{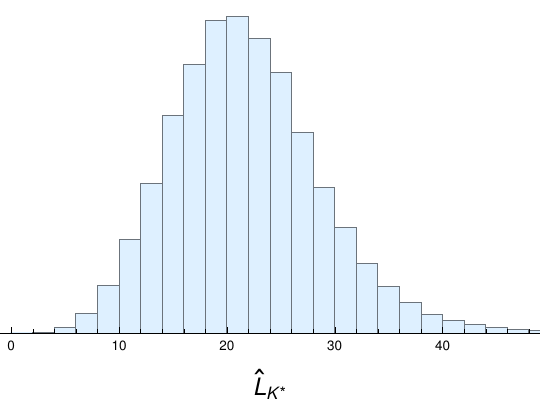} 
			\includegraphics[width=7cm]{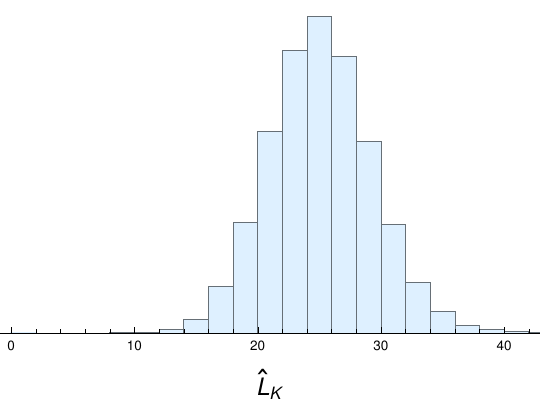} 		 
			\includegraphics[width=7cm]{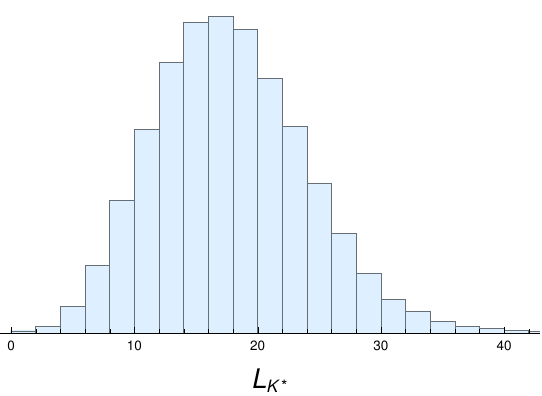}
			\includegraphics[width=7cm]{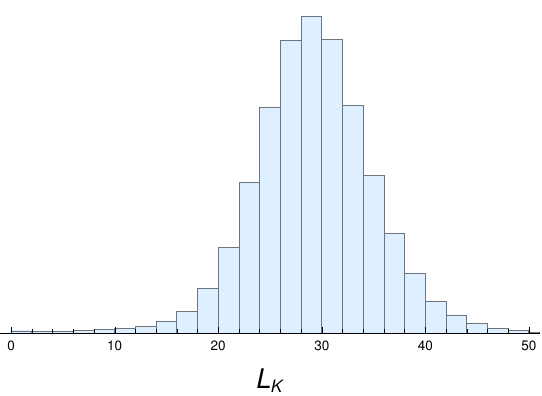} 
			\includegraphics[width=7cm]{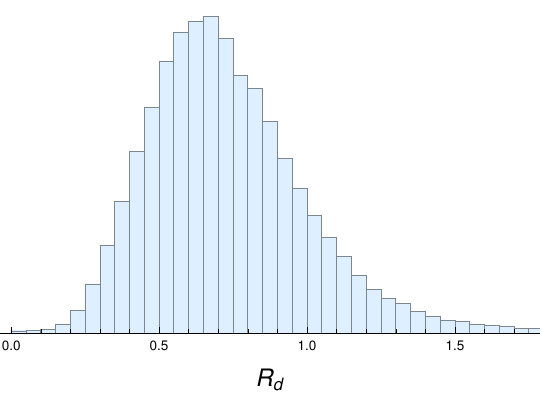} 
			\includegraphics[width=7cm]{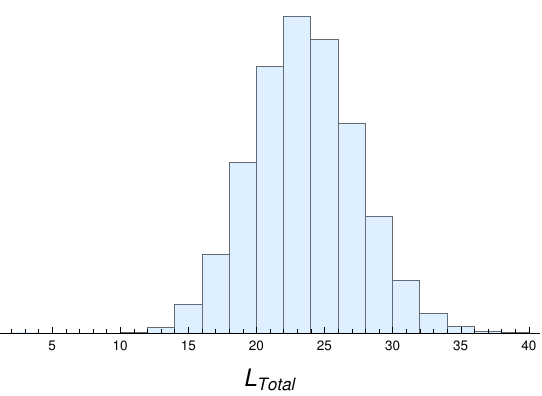} 
		\end{center}
		\caption{The SM distributions for the observables discussed in secs.~\ref{sec:th-VV}, \ref{sec:th-PP} and \ref{sec:th-PV-VP}.}
		\label{fig:obs_dist}
	\end{figure}

	We provide the SM and experimental central values along with the corresponding $1\sigma$ uncertainties for all the observables discussed in secs.~\ref{sec:th-VV}, \ref{sec:th-PP} and \ref{sec:th-PV-VP}. in table~\ref{tab:obs_val}.
	\begin{table}[ht]
		\centering
		\begin{tabular}{|c|c|c|}\hline
			\textbf{Observable} & \textbf{SM} & \textbf{Experiment}\\\hline
			$L_{K^*\bar{K}^*}$&$19.53^{+9.14}_{-6.64}$&$4.43\pm 0.92$\\
			$L_{K\bar{K}}$&$26.00^{+3.88}_{-3.59}$&$14.58\pm3.37$\\
			$\hat{L}_{K^*}$&$21.30^{+7.19}_{-6.30}$&$-$\\
			$\hat{L}_{K}$&$25.01^{+4.21}_{-4.07}$&$-$	\\
			$L_{K^*}$&$17.44^{+6.59}_{-5.82}$&$-$\\
			$L_{K}$&$29.16^{+5.49}_{-5.25}$&$-$\\
			$R_d$&$0.70^{+0.30}_{-0.22}$&$-$\\
			$L_{Total}$&$23.48^{+3.95}_{-3.82}$&$-$\\\hline				
		\end{tabular}
		\caption{Experimental values of the different observables discussed in discussed in secs.~\ref{sec:th-VV}, \ref{sec:th-PP} and \ref{sec:th-PV-VP}. Experimental values exist only for $L_{K^*\bar{K}^*}$ and $L_{K\bar{K}}$.}
		\label{tab:obs_val}
	\end{table}		
	The SM distributions for the observables discussed in sec.~\ref{sec:th-PV-VP} are provided in fig.~\ref{fig:obs_dist}.
	A few comments regarding the nature of the SM estimates and distributions for the $L_{K\bar{K}}$ and the $L_{K^*\bar{K}^*}$ observables are in order. One can see from table~\ref{tab:obs_val} that the uncertainties for $L_{K^*\bar{K}^*}$ are much larger than those for $L_{K\bar{K}}$. It is also evident from fig.~\ref{fig:obs_dist} that the former is also much more asymmetric than the later. The source of both these effects can be traced back to the form factors. The Lattice HPQCD collaboration, for the first time, has been able to measure the $B\to K$ form factors fover the entire relevant $q^2$ range. As such, the uncertainties on these form factors are much larger than the uncertainties on the $B\to K^*$ form factors, and this is precisely the reason behind $L_{K\bar K}$ being more symmetric and less uncertain in comparison to $L_{K^*\bar{K}^*}$.
	
	The experimental numbers $L_{K^*\bar{K}^*}$ and $L_{K\bar{K}}$ exhibit a deviation of $2.6\sigma$ and $2.4\sigma$ respectively from their corresponding SM values respectively. This is indeed interesting, and may be hinting towards the presence of NP. 
	
	\section{NP dependence} \label{sec:NP}
	In accordance to the current phenomenological implications of the $b\to sll$ observables, we start by assuming NP effects in the $b\to s$ sector alone. The relevant hamltonian for a general $b\to q$ transition at the scale $m_b$ is provided in appendix A.1 of  ref.~\cite{Biswas:2023pyw}. The Wilson Coefficients (WC's) relevant for the simultaneous explanation of both the $L_{K^{(*)}\bar{K}^{(*)}}$ observables are $C_{1s}^{NP}$, $C_{4s}^{NP}$ and $C_{8g,s}^{NP}$. The NP WC $C_{6s}^{NP}$ although capable of explaining $L_{K\bar{K}}$, cannot explain $L_{K^*\bar{K}^*}$. This is because of the fact that the dependence of $L_{K\bar{K}}$ on the linear term in $C_{6s}^{NP}$ is much starker (about 12 times) as compared to that of $L_{K^*\bar{K}^*}$\footnote{The complete dependence of these observables on the coefficients $C_{4s}^{NP}$, $C_{6s}^{NP}$ and $C_{8g,s}^{NP}$ are provided in eqns.4.1 and 4.2 of ref.~\cite{Biswas:2023pyw}}. The value of the coefficient $C_{1s}^{NP}$ required for a simultaneous explanation of both these observables is about 60\% of its SM value, and is discarded by the constraints discussed in ref.~\cite{Lenz:2019lvd}.
	\begin{figure}[ht]
		\centering
		\includegraphics[width=7.5cm]{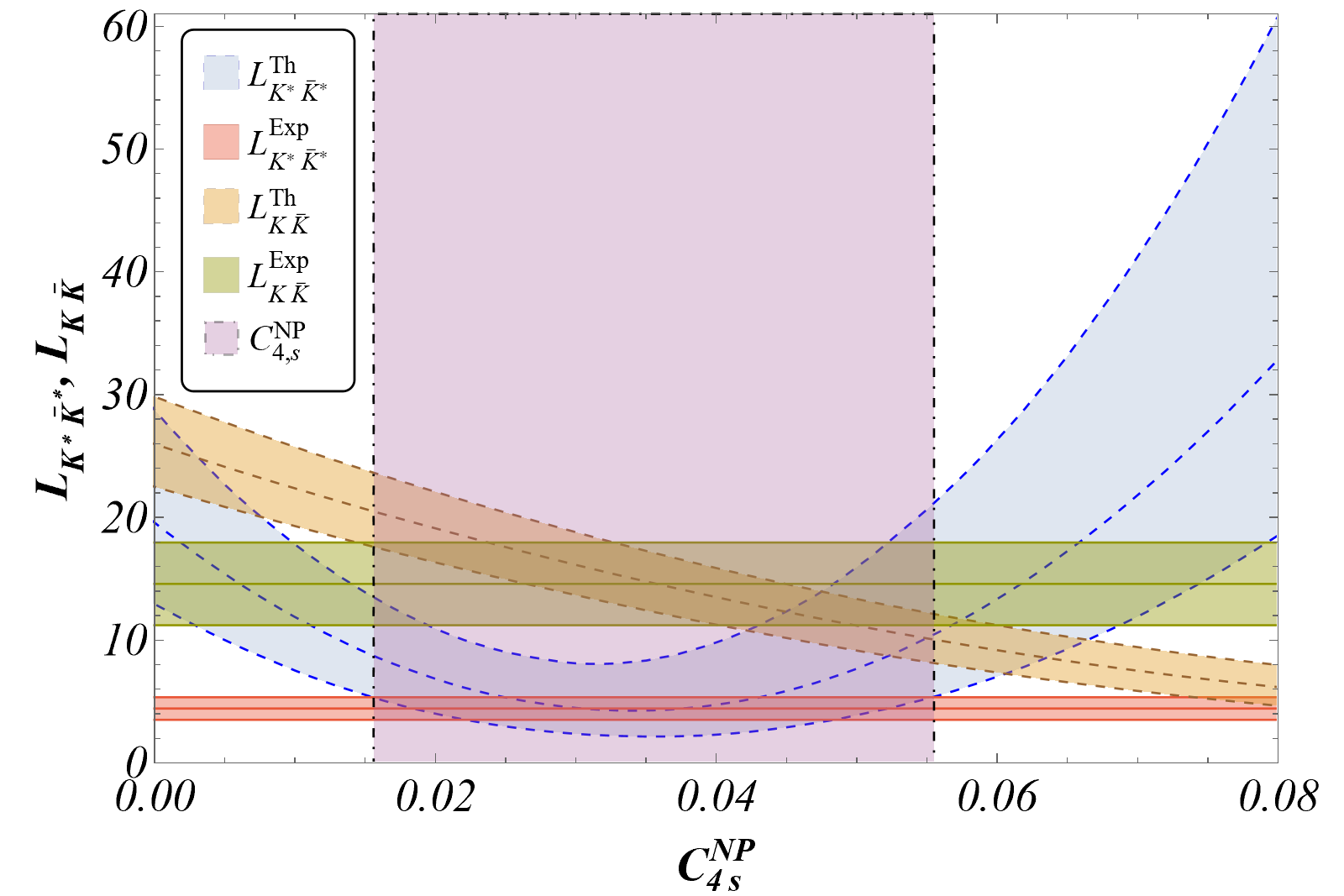}
		\includegraphics[width=7.5cm]{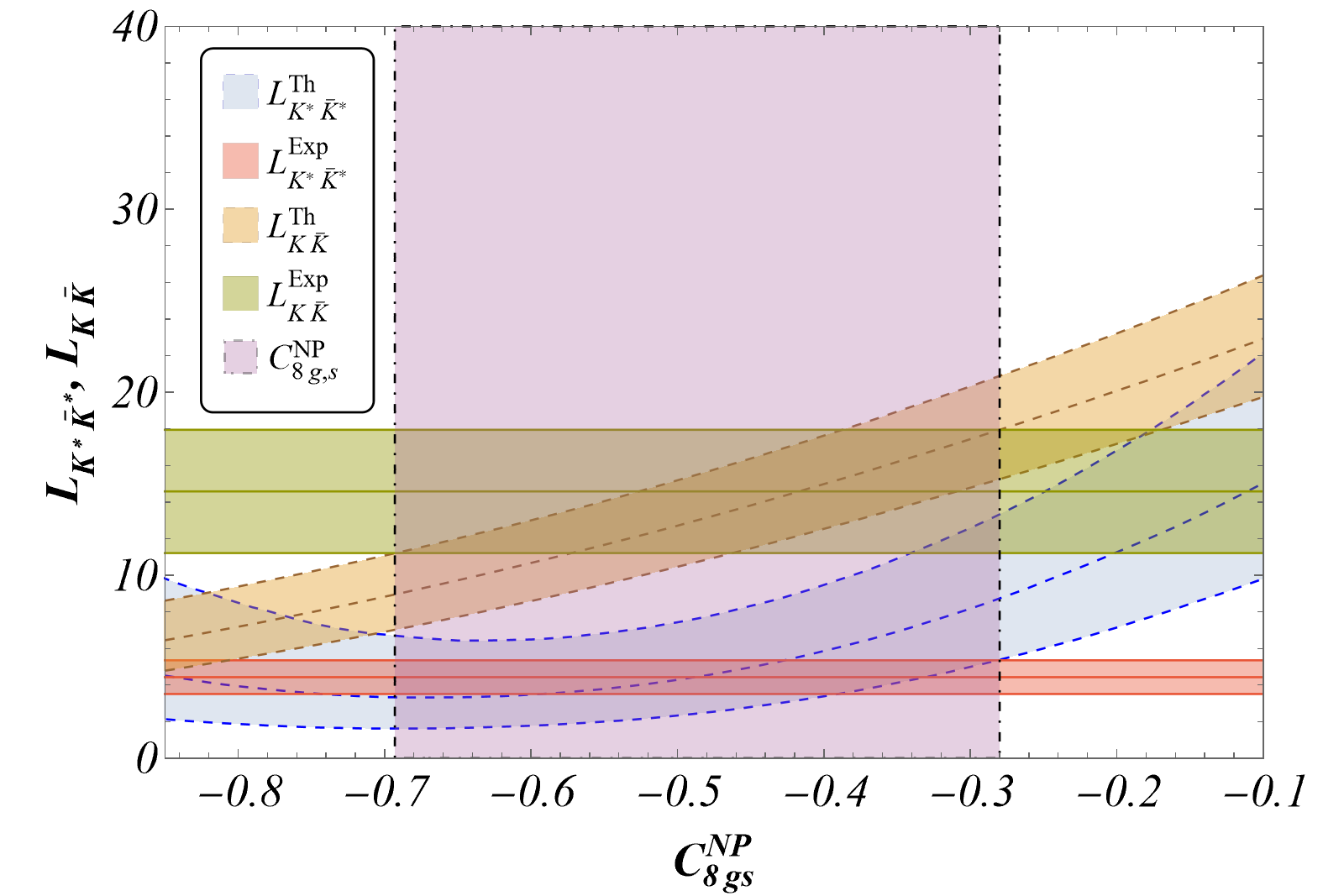}
		\caption{ Variation of $L_{K^*\bar{K}^*}$ and $L_{K\bar{K}}$ w.r.t ${\cal C}_{4s}^{\rm NP}$ (on the left) and ${\cal C}_{8g,s}^{\rm NP}$ on the right. The region relevant for a simultaneous explanation of both these observables is specified using magenta in both the figures.}
		\label{fig:c4s_c8gs_range}
	\end{figure}
	Hence the NP WC's (taken one at a time) with the potential to explain both the observables simultaneously are $C_{4s}^{NP}$ and $C_{8g,s}^{NP}$. The corresponding ranges for both thes WC's are displayed in fig.~\ref{fig:c4s_c8gs_range}.
	
	However a further careful comparison of the experimental values of the BR's instrumental in constructing these observables with their corresponding SM estimates point towards the possibility that NP might not, in this case, be affecting only the $b\to s$ sector but the $b\to d$ sector as well. In order to explain why, we provide the values of the corresponding BR's in tables~\ref{tab:BrPP} and~\ref{tab:BrVV} for the observables corresponding to the PP and the VV cases respectively.
	\begin{table}[h]
		\begin{center}
			\tabcolsep=1.3cm\begin{tabular}{|c|c|}
				\hline\multicolumn{2}{|c|}{
					$BR(\bar{B}_d\rightarrow K^0 \bar{K}^0 )$ $[10^{-6}]$}  \\ \hline
				SM (QCDF)& Experiment \\
				\hline
				$1.09^{+0.29}_{-0.20}$&$1.21\pm 0.16$ \cite{Workman:2022ynf,Belle:2012dmz,BaBar:2006enb}\\
				\hline
			\end{tabular}
			\vskip 1pt
			\tabcolsep=1.15cm\begin{tabular}{|c|c|}
				\hline\multicolumn{2}{|c|}{
					$BR(\bar{B}_s\rightarrow K^0 \bar{K}^0  )$ $[10^{-5}]$}  \\ \hline
				SM (QCDF)& Experiment \\
				\hline
				$2.80^{+0.89}_{-0.62}$&$1.76\pm 0.33$~\cite{Workman:2022ynf, LHCb:2020wrt,Belle:2015gho}\\
				\hline
			\end{tabular}
		\end{center}
		\caption{Branching ratios for pseudoscalar-pseudoscalar final states. A 7\% relative uncertainty is added in quadrature for the $B_s$ decay due to $B_s$-mixing.}
		\label{tab:BrPP}
	\end{table}
	
\begin{figure}[h]
		\centering
		\includegraphics[width=0.7\textwidth]{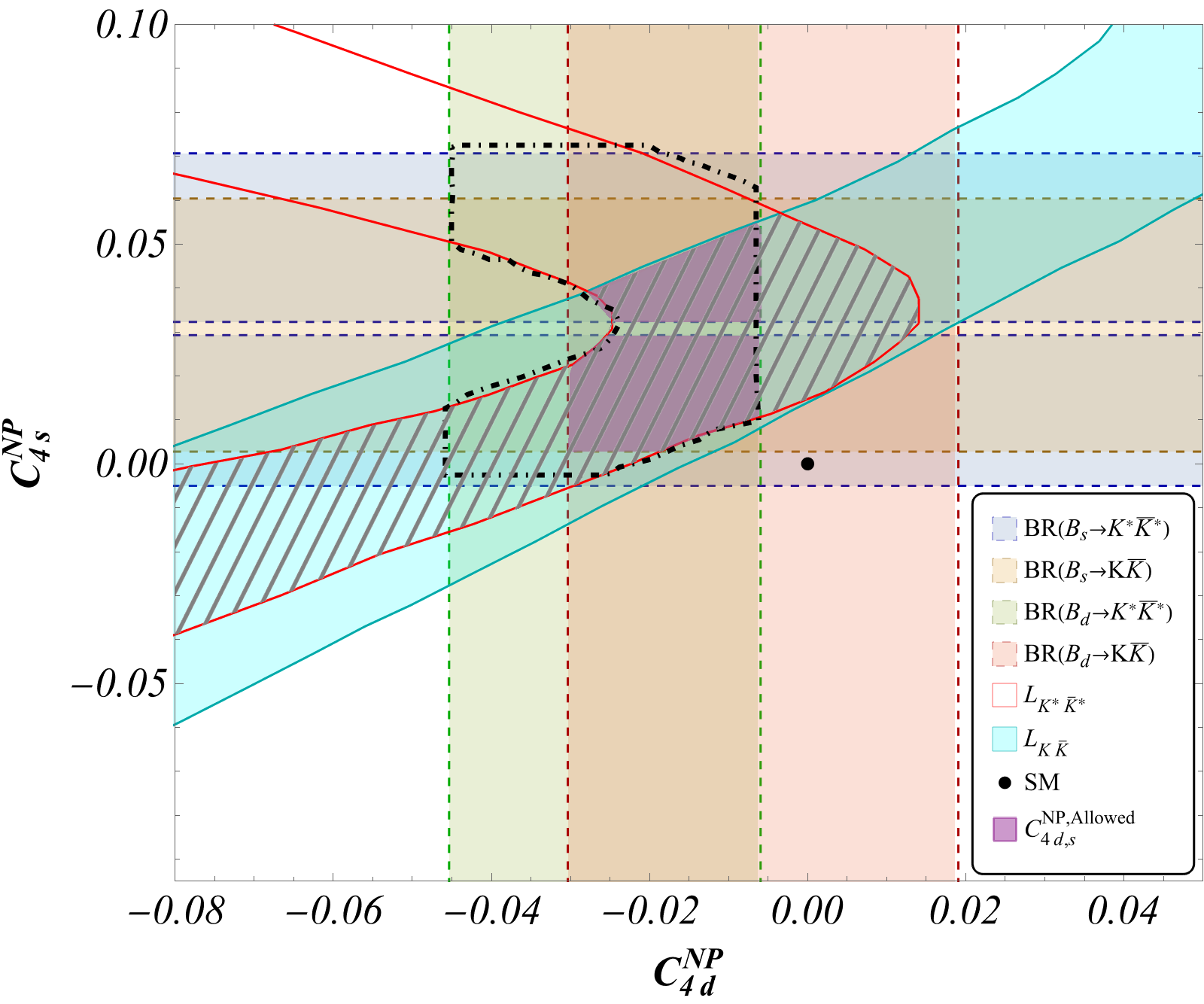}
		\caption{
			Allowed region for ${\cal C}_{4d}^{\rm NP}$-${\cal C}_{4s}^{\rm NP}$ accommodating the constraints from the measured $L$ observables and individual branching ratios, fixing ${\cal C}_{6d,6s}^{\rm NP}=0$ (magenta region) and letting ${\cal C}_{6d,6s}$ float freely (enlarged region delimited by a black dot-dashed line). We used the full expression of the $L$ observables. Notice that the enlarged region does not expand closer to the SM point. The hatched region represents the values allowed by the two measured $L$ observables only.}
		\label{fig:BR1}
	\end{figure}

\begin{figure}[h]
		\centering
		\includegraphics[width=0.7\textwidth]{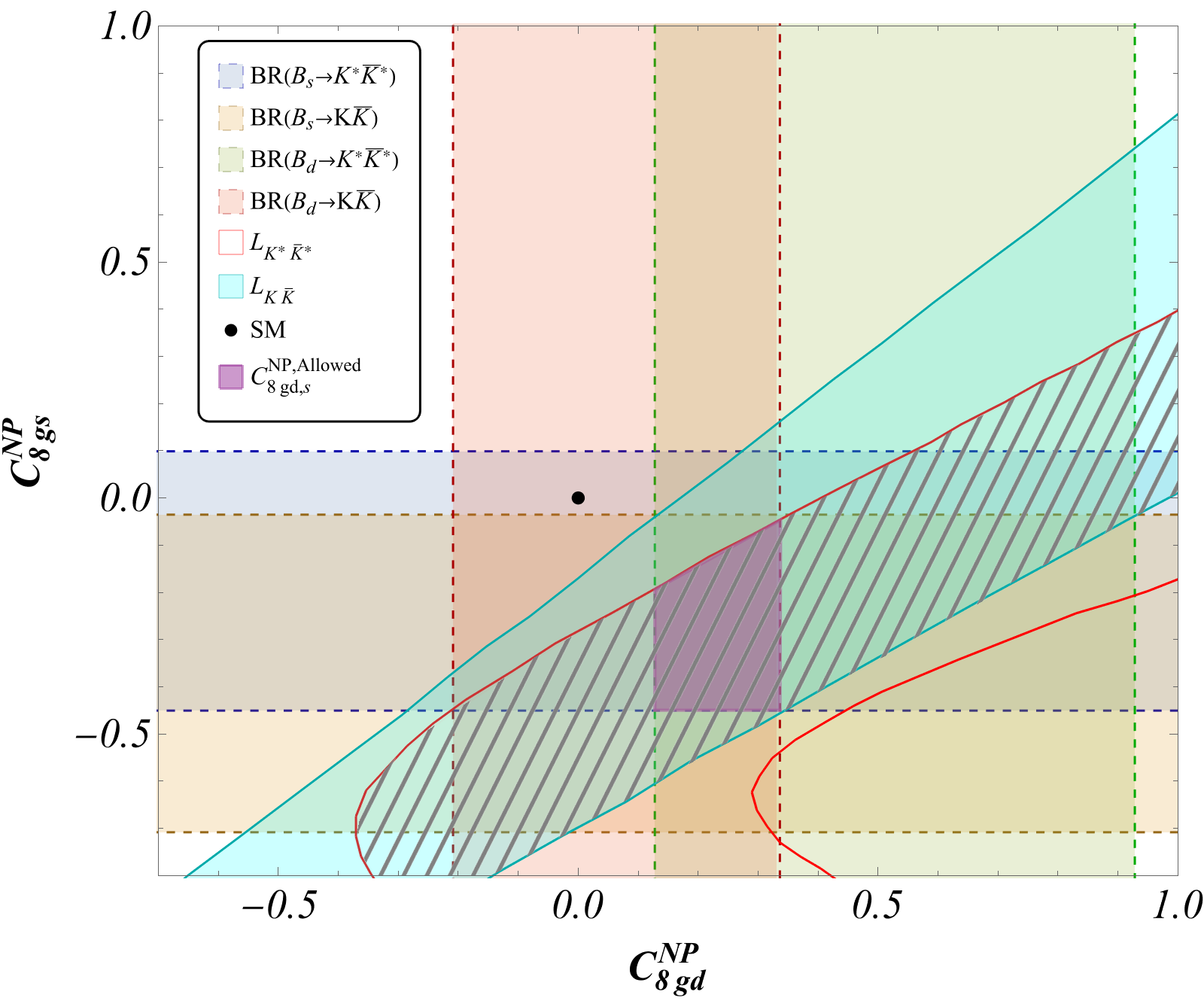}
		\caption{
			Allowed region for NP contributions to ${\cal C}_{8gd}^{\rm NP}$-${\cal C}_{8gs}^{\rm NP}$ accommodating the constraints from the measured $L$ observables and individual branching ratios (magenta region). We used the full expression of the $L$ observables. The hatched region represents the values allowed by the two measured $L$ observables only.}
		\label{fig:BR3}
	\end{figure}

	\begin{table}[t]
		\begin{center}
			\tabcolsep=2.33cm\begin{tabular}{|c|c|}
				\hline\multicolumn{2}{|c|}{Longitudinal
					$BR(\bar{B}_d\rightarrow K^{*0} \bar{K}^{*0})$ $[10^{-7}]$ }  \\ 
				\hline
				SM (QCDF)& Experiment\\
				\hline
				$2.27^{+0.98}_{-0.74}$&$6.04^{+1.81}_{-1.78}$\\
				\hline
			\end{tabular}
			
			\vskip 1pt
			
			\tabcolsep=2.33cm\begin{tabular}{|c|c|}
				\hline\multicolumn{2}{|c|}{Longitudinal
					$BR(\bar{B}_s\rightarrow K^{*0}\bar{K}^{*0} )$ $[10^{-6}]$ }  \\ 
				\hline
				SM (QCDF)& Experiment\\
				\hline
				$4.36^{+2.23}_{-1.65}$&$2.62^{+0.85}_{-0.75}$\\
				\hline
			\end{tabular}
			\caption{Branching ratios for Vector-Vector final states. A 7\% relative uncertainty is added in quadrature due to $B_s$-mixing for the $B_s$ decay. The details regarding the experimental estimates for the longitudinal BR's can be found in ref.~\cite{Biswas:2023pyw} and the references therein. 
			}
			\label{tab:BrVV}
		\end{center}
	\end{table}
	A closer look at these tables reveals that 	the SM estimate $BR(\bar{B}_s\rightarrow K^{*0}\bar{K}^{*0} )$ in particular is about $1.8\sigma$ away from the corresponding experimental numbers. This provides a legitimate motivation for exploring the scope of simultaneous NP in both $b\to s,d$ transitions as an explanation to these observables, along with the BR's instrumental for constructing them. The region in the parameter spaces corresponding to $C_{4s,d}^{NP}$ and $C_{8g,s,d}^{NP}$ that simultaneously explain all the BR's and the L observables are displayed in figs.~\ref{fig:BR1} and~\ref{fig:BR3}.
	
	\section{Disentangling NP} \label{sec:pattern}
	
\begin{figure}[h]
		\centering
		\includegraphics[width=0.8\textwidth]{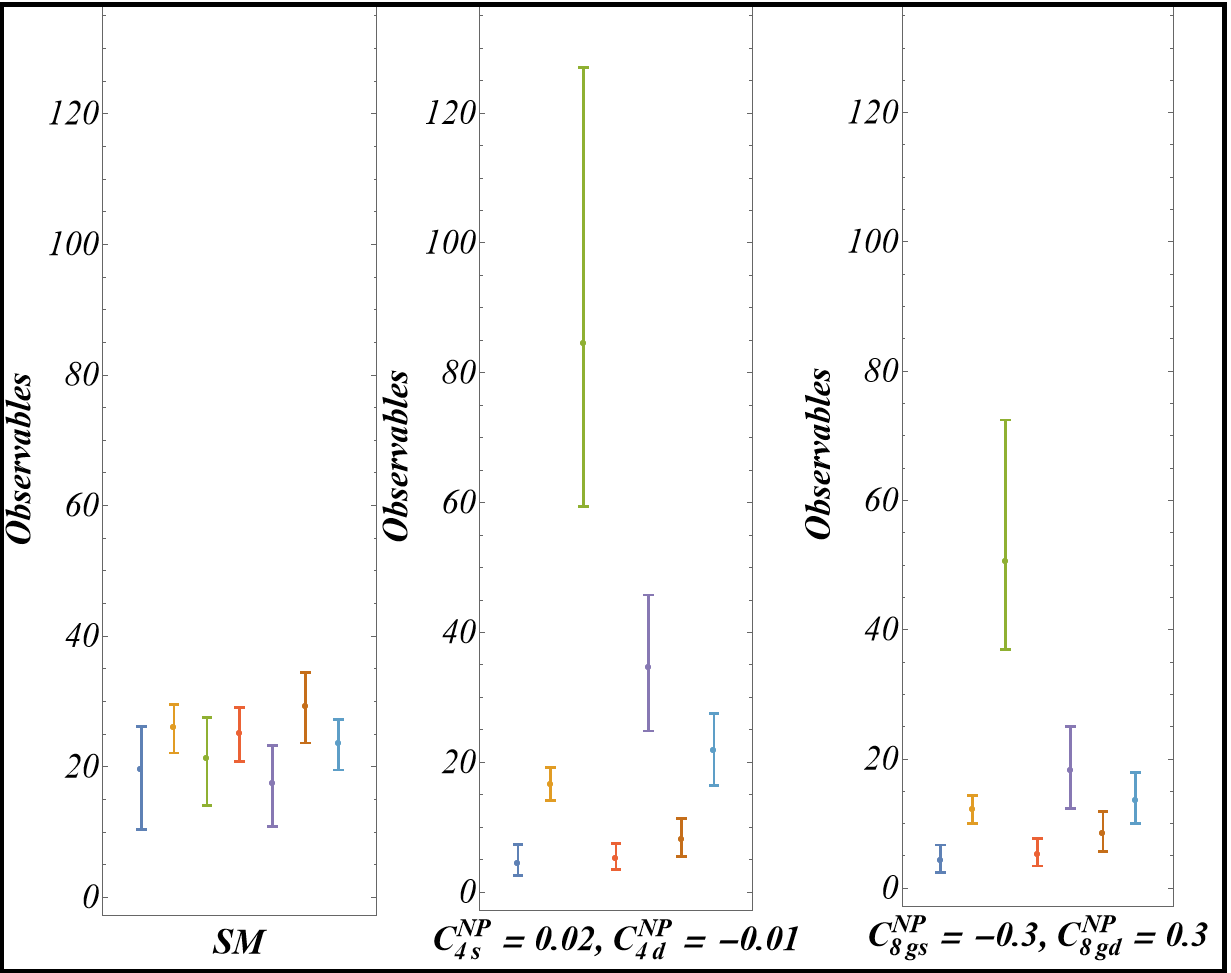}\qquad\\
		\centering
		\includegraphics[width=0.4\textwidth]{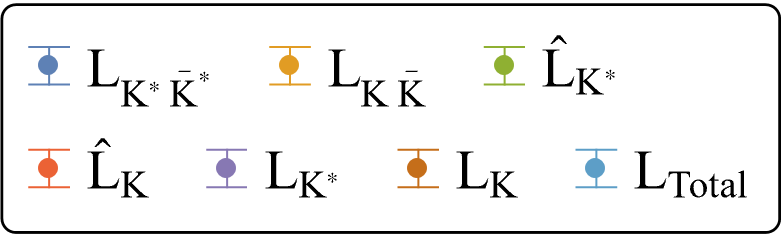}
		\caption{Predictions within the SM and different scenarios at specific NP points illustrating the patterns to be expected in each case, 
			assuming NP enters both $b\to s$ and $b\to d$ transitions. The specific benchmark values for the NP Wilson Coefficients are taken in agreement with the magenta regions shown in figs.~\ref{fig:BR1} and \ref{fig:BR3}.}
		\label{fig:dev-patternsII}
	\end{figure}

	Finally, we consider the NP dependencies of the observables defined in sec.~\ref{sec:th-PV-VP} which include the mixed modes. One can easily verify from table~\ref{tab:obs_val} that $\hat{L}_{K^{(*)}}$, $L_{K^{(*)}}$ and $L_{Total}$ are all consistent with each other as well as with $L_{K^{(*)}K^{(*)}}$ as far as their SM estimates are concerned. Tagged measurements for $BR(B_{d,s}\to K^{*0}\bar{K}^0, \bar{K}^{*0}K^0)$ are non-existent in the literature thus far. As far as the untagged measurements are concerned, experimental numbers do exist for the BR corresponding to the $B_s$ decay~\cite{LHCb:2019vww} but for the $B_d$ decay there is only an upper limit~\cite{LHCb:2015oyu}. As such, we study the NP dependencies of the mixed observables on the relevant WC's which can potentially explain $L_{K^{(*)}K^{(*)}}$ together with the corresponding BR's simultaneously (i.e. $C_{4s,d}^{NP}$ and $C_{8g,s,d}^{NP}$). Figure~\ref{fig:dev-patternsII} displays the behaviour of these observables in the SM and for benchmark values of the NP WC's $C_{4s,d}^{NP}$ and $C_{8g,s,d}^{NP}$. Interestingly, a possible solution to the non-leptonic puzzle using a model based on an $S_1$ scalar leptoquark and a TeV-scale right-handed neutrino was proposed in ref.~\cite{Lizana:2023kei}. The model produces the required NP contribution to the Wilson coefficient of chromomagnetic operators ${\cal O}_{8gd,s}$ in agreement with all constraints, specially the very stringent $B \to X_{s,d}\gamma$ ones. A further nice property of this model is that it can also explain the LFUV anomalies in charged-current B decays and as an interesting byproduct produces an enhancement of ${\cal B}(B \to K \nu \bar{\nu})$.
	In the future, with the advent of experimental data on the mixed modes, one should be able to confidently infer whether the deviations in the $L_{K^{(*)}K^{(*)}}$ observables are indeed due to NP. If the experimental estimates of the mixed observable are indeed different from each other, the pattern of their differences will be indicative of the dominant NP contribution.
	
	\section{Conclusion} \label{sec:conclusions}
	Following the discussions presented by the authors in ref.~\cite{Alguero:2020xca} in relation to an optimized observable constructed out of $B_{s,d}\to K^{*0}\bar{K}^{*0}$ modes, we extend their work to related decays with the same quark content, but different spins for the outgoing mesons, i.e. ${\bar B}_{d,s} \to K^{0} \bar{K}^{0}$, ${\bar B}_{d,s} \to K^{0} \bar{K}^{*0}$ and ${\bar B}_{d,s} \to \bar{K}^{0} {K^{*0}}$ together with their CP conjugate partners in the present article. We have designed optimised observables for these decays, with reduced hadronic uncertainties, mainly coming from form factors and power-suppressed infrared divergences, thanks to $U$-spin symmetry and QCD factorisation. We found a 2.4 $\sigma$ deviation in the pseudoscalar-pseudoscalar mode ${\bar B}_{d,s} \to K^{0} \bar{K}^{0}$ whereas the lack of experimental information prevented us from analysing the pseudoscalar-vector modes ${\bar B}_{d,s} \to K^{0} \bar{K}^{*0}$ and ${\bar B}_{d,s} \to \bar{K}^{0} {K^{*0}}$ in more detail. 
	
	We reconsidered some NP scenarios able to explain the tensions between the SM prediction and data in $L_{K^* \bar{K}^*}$ and $L_{K\bar{K}}$, focusing first on NP in $b\to s$ transitions only. It turns out that a simultaneous explanation to these observables can be attribute to ${\cal C}_{4s}$ and ${\cal C}_{8gs}$. In addition to the ratios of branching ratios such as $L_{K^* \bar{K}^*}$ and $L_{K\bar{K}}$, we also considered individual branching ratios in the same framework. It turns out that deviations occur in ${\cal B}(\bar{B}_s\to K\bar{K})$ but also in ${\cal B}(\bar{B}_d\to K^*\bar{K}^*)$, though at a more modest level than $L_{K^* \bar{K}^*}$ and $L_{K\bar{K}}$. It thus suggests that NP scenarios with contributions to both $b\to d$ and $b\to s$ transitions should be considered. 

	We have identified domains of NP contributions to ${\cal C}_{4(d,s)}$ and ${\cal C}_{8g(d,s)}$ (see Figs.~\ref{fig:BR1} and \ref{fig:BR3}) which could accommodate all the measurements related to $K\bar{K}$ and $K^*\bar{K}^*$ (both $L$-observables and individual branching ratios) within their theoretical and experimental 1$\sigma$ ranges. Significant deviations of branching ratios in the pseudoscalar-vector modes should be observed according to different patterns of deviations associated with different NP scenarios. For such scenarios, it is particularly important to measure also the pseudoscalar-vector observables $\hat{L}_K$ and $\hat{L}_{K^*}$ and not only $L_K$ and $L_{K^*}$ which have a more limited sensitivity to NP. We kept our discussion of these scenarios at a qualitative level without trying to perform a detailed statistical analysis. Since individual branching ratios are rather sensitive to the models used to describe long-distance contributions that are $1/m_b$-suppressed within QCD factorisation, we did not attempt to perform a global fit analysis, which is left for a future work.
	
	A (future) confirmation of such consistent set of deviations in these channels would be extremely valuable. It would point towards a common origin and would provide a possible strong hint of NP in the non-leptonic sector, which would require a more elaborate statistical framework than the simple approach presented here. In any case, we hope that our analysis will thus constitute a strong incentive to study these penguin-mediated modes experimentally in more detail in the coming years.

	\bibliographystyle{JHEP}
	
	\bibliography{main.bib}

\providecommand{\href}[2]{#2}\begingroup\raggedright\begin{thebibliography}{10}

\bibitem{Alguero:2020xca}
M.~Alguer\'o, A.~Crivellin, S.~Descotes-Genon, J.~Matias and M.~Novoa-Brunet,
  \emph{{A new $B$-flavour anomaly in $B_{d,s}\to K^{*0}\bar{K}^{*0}$: anatomy
  and interpretation}},
  \href{https://doi.org/10.1007/JHEP04(2021)066}{\emph{JHEP} {\bfseries 04}
  (2021) 066} [\href{https://arxiv.org/abs/2011.07867}{{\ttfamily
  2011.07867}}].

\bibitem{Amhis:2022hpm}
Y.~Amhis, Y.~Grossman and Y.~Nir, \emph{{The branching fraction of $B_s\to
  K^0\bar K^0$: Three puzzles}},
  \href{https://arxiv.org/abs/2212.03874}{{\ttfamily 2212.03874}}.

\bibitem{Li:2022mtc}
Y.~Li, G.-H.~Zhao, Y.-J.~Sun and Z.-T.~Zou, \emph{{Family Non-universal
  $Z^\prime$ Effects on $B_{d,s} \to K^{*0} {\overline K^{*0}}$ Decays in
  Perturbative QCD Approach}},
  \href{https://doi.org/10.1103/PhysRevD.106.093009}{\emph{Phys. Rev. D}
  {\bfseries 106} (2022) 093009}
  [\href{https://arxiv.org/abs/2209.13389}{{\ttfamily 2209.13389}}].

\bibitem{Fleischer:1999zi}
R.~Fleischer, \emph{{Extracting CKM phases from angular distributions of B(d,s)
  decays into admixtures of CP eigenstates}},
  \href{https://doi.org/10.1103/PhysRevD.60.073008}{\emph{Phys. Rev. D}
  {\bfseries 60} (1999) 073008}
  [\href{https://arxiv.org/abs/hep-ph/9903540}{{\ttfamily hep-ph/9903540}}].

\bibitem{Fleischer:1999pa}
R.~Fleischer, \emph{{New strategies to extract Beta and gamma from B(d)
  ---\ensuremath{>} pi+ pi- and B(S) ---\ensuremath{>} K+ K-}},
  \href{https://doi.org/10.1016/S0370-2693(99)00640-1}{\emph{Phys. Lett. B}
  {\bfseries 459} (1999) 306}
  [\href{https://arxiv.org/abs/hep-ph/9903456}{{\ttfamily hep-ph/9903456}}].

\bibitem{Fleischer:2007hj}
R.~Fleischer, \emph{{$B_{s,d} \to \pi \pi, \pi K, KK$: Status and Prospects}},
  \href{https://doi.org/10.1140/epjc/s10052-007-0391-7}{\emph{Eur. Phys. J. C}
  {\bfseries 52} (2007) 267} [\href{https://arxiv.org/abs/0705.1121}{{\ttfamily
  0705.1121}}].

\bibitem{Fleischer:2010ib}
R.~Fleischer and R.~Knegjens, \emph{{In Pursuit of New Physics With $B^0_s \to
  K^+K^-$}}, \href{https://doi.org/10.1140/epjc/s10052-010-1532-y}{\emph{Eur.
  Phys. J. C} {\bfseries 71} (2011) 1532}
  [\href{https://arxiv.org/abs/1011.1096}{{\ttfamily 1011.1096}}].

\bibitem{Fleischer:2016jbf}
R.~Fleischer, R.~Jaarsma and K.K.~Vos, \emph{{New strategy to explore CP
  violation with $B^0_s\to K^-K^+$}},
  \href{https://doi.org/10.1103/PhysRevD.94.113014}{\emph{Phys. Rev. D}
  {\bfseries 94} (2016) 113014}
  [\href{https://arxiv.org/abs/1608.00901}{{\ttfamily 1608.00901}}].

\bibitem{Bhattacharya:2022akr}
B.~Bhattacharya, S.~Kumbhakar, D.~London and N.~Payot, \emph{{A U-spin Puzzle
  in $B$ Decays}},  \href{https://arxiv.org/abs/2211.06994}{{\ttfamily
  2211.06994}}.

\bibitem{Beneke:2000ry}
M.~Beneke, G.~Buchalla, M.~Neubert and C.T.~Sachrajda, \emph{{QCD factorization
  for exclusive, nonleptonic B meson decays: General arguments and the case of
  heavy light final states}},
  \href{https://doi.org/10.1016/S0550-3213(00)00559-9}{\emph{Nucl. Phys. B}
  {\bfseries 591} (2000) 313}
  [\href{https://arxiv.org/abs/hep-ph/0006124}{{\ttfamily hep-ph/0006124}}].

\bibitem{Beneke:2001ev}
M.~Beneke, G.~Buchalla, M.~Neubert and C.T.~Sachrajda, \emph{{QCD factorization
  in B ---\ensuremath{>} pi K, pi pi decays and extraction of Wolfenstein
  parameters}},
  \href{https://doi.org/10.1016/S0550-3213(01)00251-6}{\emph{Nucl. Phys. B}
  {\bfseries 606} (2001) 245}
  [\href{https://arxiv.org/abs/hep-ph/0104110}{{\ttfamily hep-ph/0104110}}].

\bibitem{Beneke:2003zv}
M.~Beneke and M.~Neubert, \emph{{QCD factorization for B ---\ensuremath{>} PP
  and B ---\ensuremath{>} PV decays}},
  \href{https://doi.org/10.1016/j.nuclphysb.2003.09.026}{\emph{Nucl. Phys. B}
  {\bfseries 675} (2003) 333}
  [\href{https://arxiv.org/abs/hep-ph/0308039}{{\ttfamily hep-ph/0308039}}].

\bibitem{Beneke:2006hg}
M.~Beneke, J.~Rohrer and D.~Yang, \emph{{Branching fractions, polarisation and
  asymmetries of B ---\ensuremath{>} VV decays}},
  \href{https://doi.org/10.1016/j.nuclphysb.2007.03.020}{\emph{Nucl. Phys. B}
  {\bfseries 774} (2007) 64}
  [\href{https://arxiv.org/abs/hep-ph/0612290}{{\ttfamily hep-ph/0612290}}].

\bibitem{Bartsch:2008ps}
M.~Bartsch, G.~Buchalla and C.~Kraus, \emph{{B ---\ensuremath{>} V(L) V(L)
  Decays at Next-to-Leading Order in QCD}},
  \href{https://arxiv.org/abs/0810.0249}{{\ttfamily 0810.0249}}.

\bibitem{Descotes-Genon:2006spp}
S.~Descotes-Genon, J.~Matias and J.~Virto, \emph{{Exploring B(d,s)
  ---\ensuremath{>} KK decays through flavour symmetries and
  QCD-factorisation}},
  \href{https://doi.org/10.1103/PhysRevLett.97.061801}{\emph{Phys. Rev. Lett.}
  {\bfseries 97} (2006) 061801}
  [\href{https://arxiv.org/abs/hep-ph/0603239}{{\ttfamily hep-ph/0603239}}].

\bibitem{Descotes-Genon:2007iri}
S.~Descotes-Genon, J.~Matias and J.~Virto, \emph{{Penguin-mediated $B_{d,s} \to
  VV$ decays and the $B_s - \bar{B}_s$ mixing angle}},
  \href{https://doi.org/10.1103/PhysRevD.76.074005}{\emph{Phys. Rev. D}
  {\bfseries 76} (2007) 074005}
  [\href{https://arxiv.org/abs/0705.0477}{{\ttfamily 0705.0477}}].

\bibitem{Descotes-Genon:2011rgs}
S.~Descotes-Genon, J.~Matias and J.~Virto, \emph{{An analysis of $B_{d,s}$
  mixing angles in presence of New Physics and an update of $B_s
  \to\bar{K}^{0*} anti-K^{0*}$}},
  \href{https://doi.org/10.1103/PhysRevD.85.034010}{\emph{Phys. Rev. D}
  {\bfseries 85} (2012) 034010}
  [\href{https://arxiv.org/abs/1111.4882}{{\ttfamily 1111.4882}}].

\bibitem{Biswas:2023pyw}
A.~Biswas, S.~Descotes-Genon, J.~Matias and G.~Tetlalmatzi-Xolocotzi, \emph{{A
  new puzzle in non-leptonic B decays}},
  \href{https://doi.org/10.1007/JHEP06(2023)108}{\emph{JHEP} {\bfseries 06}
  (2023) 108} [\href{https://arxiv.org/abs/2301.10542}{{\ttfamily
  2301.10542}}].

\bibitem{Fazzini:2018dyq}
{\scshape LHCb} collaboration, \emph{{Flavour Tagging in the LHCb experiment}},
  \href{https://doi.org/10.22323/1.321.0230}{\emph{PoS} {\bfseries LHCP2018}
  (2018) 230}.

\bibitem{Lenz:2019lvd}
A.~Lenz and G.~Tetlalmatzi-Xolocotzi, \emph{{Model-independent bounds on new
  physics effects in non-leptonic tree-level decays of B-mesons}},
  \href{https://doi.org/10.1007/JHEP07(2020)177}{\emph{JHEP} {\bfseries 07}
  (2020) 177} [\href{https://arxiv.org/abs/1912.07621}{{\ttfamily
  1912.07621}}].

\bibitem{Workman:2022ynf}
{\scshape Particle Data Group} collaboration, \emph{{Review of Particle
  Physics}}, \href{https://doi.org/10.1093/ptep/ptac097}{\emph{PTEP} {\bfseries
  2022} (2022) 083C01}.

\bibitem{Belle:2012dmz}
{\scshape Belle} collaboration, \emph{{Measurements of branching fractions and
  direct CP asymmetries for B\textrightarrow{}K\ensuremath{\pi},
  B\textrightarrow{}\ensuremath{\pi}\ensuremath{\pi} and B\textrightarrow{}KK
  decays}}, \href{https://doi.org/10.1103/PhysRevD.87.031103}{\emph{Phys. Rev.
  D} {\bfseries 87} (2013) 031103}
  [\href{https://arxiv.org/abs/1210.1348}{{\ttfamily 1210.1348}}].

\bibitem{BaBar:2006enb}
{\scshape BaBar} collaboration, \emph{{Observation of $B^{+} \to \bar{K}^0
  K^{+}$ and $B^0 \to K^0 \bar{K}^0$}},
  \href{https://doi.org/10.1103/PhysRevLett.97.171805}{\emph{Phys. Rev. Lett.}
  {\bfseries 97} (2006) 171805}
  [\href{https://arxiv.org/abs/hep-ex/0608036}{{\ttfamily hep-ex/0608036}}].

\bibitem{LHCb:2020wrt}
{\scshape LHCb} collaboration, \emph{{Measurement of the branching fraction of
  the decay $B_s^0\to K_S^0 K_S^0$}},
  \href{https://doi.org/10.1103/PhysRevD.102.012011}{\emph{Phys. Rev. D}
  {\bfseries 102} (2020) 012011}
  [\href{https://arxiv.org/abs/2002.08229}{{\ttfamily 2002.08229}}].

\bibitem{Belle:2015gho}
{\scshape Belle} collaboration, \emph{{Observation of the decay
  $B_s^0\rightarrow K^0\overline{K}^0$}},
  \href{https://doi.org/10.1103/PhysRevLett.116.161801}{\emph{Phys. Rev. Lett.}
  {\bfseries 116} (2016) 161801}
  [\href{https://arxiv.org/abs/1512.02145}{{\ttfamily 1512.02145}}].

\bibitem{LHCb:2019vww}
{\scshape LHCb} collaboration, \emph{{Amplitude analysis of $B^{0}_{s}
  \rightarrow K^{0}_{\textrm{S}} K^{\pm}\pi^{\mp}$ decays}},
  \href{https://doi.org/10.1007/JHEP06(2019)114}{\emph{JHEP} {\bfseries 06}
  (2019) 114} [\href{https://arxiv.org/abs/1902.07955}{{\ttfamily
  1902.07955}}].

\bibitem{LHCb:2015oyu}
{\scshape LHCb} collaboration, \emph{{First observation of the decay $B_{s}^{0}
  \to K_{S}^{0} K^{*}(892)^{0}$ at LHCb}},
  \href{https://doi.org/10.1007/JHEP01(2016)012}{\emph{JHEP} {\bfseries 01}
  (2016) 012} [\href{https://arxiv.org/abs/1506.08634}{{\ttfamily
  1506.08634}}].

\bibitem{Lizana:2023kei}
J.M.~Lizana, J.~Matias and B.A.~Stefanek, \emph{{Explaining the
  $B_{d,s}\rightarrow {K^{(*)}\bar K^{(*)}}$ non-leptonic puzzle and
  charged-current $B$-anomalies via scalar leptoquarks}},
  \href{https://arxiv.org/abs/2306.09178}{{\ttfamily 2306.09178}}.

\end{thebibliography}\endgroup
	
	%\begin{thebibliography}{99}
	%\bibitem{...}
	%....
	
	%\end{thebibliography}
	
\end{document}